# An Interactive Web Application for School-Based Physical Fitness Testing in California: Geospatial Analysis and Custom Mapping

Yawen Guo, MISM[1], Kaiyuan Hu[1], Di Hu, MS[1], Kai Zheng, PhD[1], Dan M Cooper, MD[1]
[1] University of California, Irvine, Irvine, CA, USA

**Abstract**

*Physical activity is crucial for children's healthy growth and development. In the US, most states have physical education standards. California implemented the mandated School-based Physical Fitness Testing (SB-PFT) over two decades ago. Despite the substantial effort in collecting the SB-PFT data, its research reuse has been limited due to the lack of readily accessible analytical tools. We developed a web application utilizing GeoServer, ArcGIS, and AWS to visualize the SB-PFT data. Education administrators and policymakers can leverage this user-friendly platform to gain insights into children's physical fitness trend, and identify schools and districts with successful programs to gauge the success of new physical education programs. The application also includes a custom mapping tool that allows users to compare external datasets with SB-PFT. We conclude that by incorporating advanced analytical capabilities through an informatics-based user-facing tool, this platform has great potential to encourage a broader engagement in enhancing children's physical fitness.*

**Introduction**

Physical fitness plays a vital role as a bellwether in childhood and adolescence that tracks across the lifespan[1]. During 2017-2020, the prevalence of obesity was 19.7% and affected about 14.7 million children and adolescents aged 2-19 years[2]. While there are many contributing factors such as genetics, environment, and behavior[3], childhood obesity is often accompanied by lower levels of physical activity and reduced cardiorespiratory and metabolic fitness, increasing the risk of developing cardiovascular disease[4,5], type 2 diabetes[6], dyslipidemia, hypertension, certain malignancies, and nonalcoholic fatty liver disease [7,8,9].

School-based physical fitness testing(SB-PFT) is mandated in 16 states to monitor and improve children's health and fitness levels, and plays a critical role for early detection of health problems and intervention strategies[10]. California initiated its statewide SB-PFT in 1995; test result data since 1999 is publicly available in the 5th, 7th, and 9th grades. Existing research on SB-PFT has spanned from broad national analyses to focused state-level investigations. An investigation of teacher's use of fitness tests, involving 325 physical education teachers across 10 states, highlighted the isolated role of fitness testing and identified the need for consistency and reproducibility of SB-PFT across states[11]. Existing research also determined the association between SB-PFT awards and meeting physical activity guidelines among children with disabilities[12], evaluating the association among SB-PFT and bullying, weight-based teasing, and gender discrimination[13]. Within California, fitness outcomes from SB-PFT improved from 2003 to 2008, despite a persistent rise in obesity rates among incoming 5th graders[14]. While the main purpose of the testing is to educate students about health as well as physical fitness and to motivate them to increase physical activity [15,16,17], the data collected have great potential to inform socioeconomic, racial, and ethnic disparities in children's health[10]. However, the comprehensive dataset remains underutilized due, in part, to complexities and challenges in accessibility.

Geospatial analysis and visualization[18] has been used in health sciences research such as evaluating access to care in rural areas[19], access to dental caries treatment[20], overweight and obesity in school-age children[21], and the health impacts of environmental exposures[22]. These tools offer intuitive and accessible insights into the geographical distribution of health-related metrics and help discover patterns, trends, and disparities in physical fitness of children. Social determinants of health(SDoH) related to environment, society, and economy influence the physical fitness and health outcomes of children[23]. The potential of integrating SB-PFT results with SDoH promises a comprehensive understanding of the challenges and possibilities for improving the fitness levels of children. To our knowledge, no research has specifically aimed at visualizing SB-PFT outcomes across the entire state of California, nor has it conducted school district-level SB-PFT analysis alongside SDoH.

To enhance the accessibility and analysis of SB-PFT data, we introduce an interactive web application designed for mapping and geospatial analysis of California's SB-PFT results from 1999 to 2019 at the school and school district level. The platform enables users to tailor their inquiries by selecting specific years, test assessments, and counties. Furthermore, its intuitive interface supports personalized map creation and allows users to incorporate their own datasets for additional visualization, analysis, and research. After a thorough reorganization and cleaning of the existing SB-PFT data, we present a dataset that is ready for use and facilitates time series analysis, regression analysis, and machine learning applications potentially with other SDoH, which offers a powerful approach to identifying and addressing the multifaceted aspects of physical fitness among school-aged children. Our initiative in map visualization aims to offer a comprehensive perspective of children's physical fitness to a wide range of stakeholders, including parents, school administrative staff, public health practitioners, and policy makers, and contribute valuable insights to inform decision-making and target interventions for a broader audience.

**Methods**

The web application utilized the publicly available SB-PFT result data from the California Department of Education (CDoE), and was developed using ArcGIS, GeoServer, and Amazon Web Services (AWS) to enable public access. We processed the research files into a readily available format suitable for in-depth data analysis tasks for users. The effectiveness of our approach was demonstrated through a case study that analyzed the relationship between students' physical fitness and selected social determinants of health (SDoH) from the American Community Survey.

***Data - PFT Statewide Research Files.*** The CDoE has mandated SB-PFT and made the test results publicly available from 1999 to 2019, known as the Physical Fitness Testing Statewide Research Files (PFTSRF)[24]. Notably, the SB-PFT was suspended during 2019–2021 school years due to the COVID-19 pandemic. During the 2018-2019 school year, according to the CDoE PFT Statewide Research Files, there were 8,311 public schools and 1,009 school districts. The numbers of schools and school districts may vary over the years due to changes in district boundaries, school openings, closures, and consolidations. The PFTSRF contains detailed annual fitness testing results at school, school district, county, and state levels, categorizing students into "Healthy Fitness Zone" (HFZ), "Needs Improvement", and "Needs Improvement - High Risk"[25]. PFTSRF reports results on 6 key test assessments: aerobic capacity, body composition, upper body strength, abdominal strength, flexibility, and trunk lift. The PFTSRF data for each year were processed to address inconsistencies, including missing values and redundant text spaces and entries. Subsequently, we created a database combining 20 years of SB-PFT results; for every school district or individual school, 6 new records were generated, with each record corresponding to a test subject. This preprocessing step ensured that the dataset is readily available for integration with the ArcGIS layer at the school district level for subsequent visualization tasks. We chose the percentage of students in the HFZ as the primary metric for visualization for its effectiveness and informativeness in providing a broad overview of physical fitness among school-aged children at the school and school district level on all 6 test assessments.

***Backend Design and Application.*** The map visualization integrated ArcGIS, GeoServer, and multiple Python libraries for effective data handling and display. The overall backend is hosted under Node.js[26] and Express.js[27], assisting in guiding the users' requests to corresponding routes. For school district level SB-PFT results, we implemented ArcGIS to merge preprocessed PFTSRF into the California school district map layer on CDScode (unique code combination of county, district, and school), and generate shapefiles (.shp)[28] linked to corresponding school districts. These shapefiles were then uploaded to GeoServer[29], an open-source Java-based application that allows users to store, view and edit geospatial data. The web application also incorporated Leaflet.js[30] as an API connecting to GeoServer to retrieve maps through Web Mapping Service (WMS) protocol and displayed on the webpage[31]. The application processes users' display selections as a HTTP POST request through middleware[32] to ensure validity, subsequently generating new HTTP GET requests to the WMS provided by GeoServer. WMS will return the requested map and dataset to the frontend for visualization. At the school level, the same design was implemented to process users' display selections.

For school level visualization, geospatial information of each school in California was retrieved from CDoE and converted to GeoJSON[33] by Pandas[34] and GeoPandas libraries for storage. After receiving a user request, a complete GeoJSON object containing both geospatial information of school and corresponding SB-PFT results was created and then fed into Mapbox GL JS API for visualization[35]. For custom map creation at school district level, the uploaded files are processed with a Python script using the Pandas, Shapely[36] and Fiona libraries to integrate with the California school district map, creating a shapefile and uploaded to GeoServer to create a new map layer.

***Web Design and AWS Deployment***. We designed the web interface through Cascading Style Sheets(CSS)[37] and Bootstrap[38] templates into five main components: an introduction page, the SB-PFT results school map, the school district map, a custom map visualization page, and an external map upload page. The Introduction page was designed to provide users with an engaging overview of the SB-PFT program and the map application, incorporating general information, references, and tutorials to enhance user familiarity and ease of use. The interface for visualizing SB-PFT results and custom datasets incorporated a form with dropdown menus and a checkbox window, offering users the flexibility to tailor their selections according to their preferences. The system allows users to select specific grades, test assessments, and school years, and geographic area of interests.

For individual school visualization, user-defined parameters along with the corresponding GeoJSON dataset were integrated into the Mapbox GL JS API, enabling the visualization of data through dynamically sized circles. These circles were designed to automatically cluster or de-cluster as users zoom in and out. Each school's percentage of students in the HFZ is visually differentiated using a color-coded scheme. For detailed information of each school, pop-up windows were made available from the circles by clicks. When users submit their selection at the school district level, a layer fetched by the backend WMS is dynamically displayed on a map provided by OpenStreetMap on the webpage through Leaflet.js, showing the percentage of students in the HFZ across selected school districts. We designed the pop-up windows to display detailed information with a legend and data range labels to provide detailed information for the visualized geographic area. For visualizing custom datasets at school district level, we designed the custom mapping interface for users to upload Excel or CSV files and create school district level mappings. The function allows users to upload one data file at a time and create multiple layers of maps. Upon the user submitting the request, the corresponding map layer is fetched from WMS and displayed.

Deployment flexibility allows the application to run both locally and on AWS, with environment-specific configurations. Locally, the application applies nodemon module[31] for live updates, while on AWS, PM2 ensures continuous service. GeoServer credentials on AWS are securely managed through AWS Secrets Manager and IAM policies, ensuring secure access and operation.

***Case Study - Integrated SB-PFT Results with SDoH Data for Explanatory Modeling.*** In this case study, we present an example to showcase the utility of our integrated SB-PFT results beyond visualization. We first merged the SB-PFT results for 5th graders with 2015-2019 data from the American Community Survey – Education Children's Tabulation by the National Center for Education Statistics[39]. By combining the two datasets, we linked children's fitness indicators to socioeconomic indexes at school district level, enabling us to conduct regression analysis to understand how socioeconomic status affects children's cardiovascular and overall fitness. We employed multiple linear regressions with 2 critical indicators of children's fitness as dependent variables: *Percentage of students in HFZ of Aerobic Capacity* and *Percentage of Students in HFZ of Body Composition*. For demonstration purposes, we selected five SDoH factors as independent variables: *the Percentage of Population Speaking a Language Other than English at Home*, *the Percentage of Noninstitutionalized Civilians with Public Health Insurance*, *the Percentage of Households with a Computer*, *the Percentage of Occupied Housing Units with No Vehicles, and the Mean Family Income (in units of 10000 dollars)*. The variables were selected based on literature demonstrating the significance of residential neighborhood and community factors in relation to children's physical activities.[40, 41, 42, 43, 44]. A total of 677 elementary or unified school districts were included in the analysis using R (Version 4.3.3). Our multiple linear regression models were assessed and met the assumptions of normality, linearity, and homoscedasticity. In addition, multicollinearity was checked through the Variance Inflation Factor (VIF), with all VIF values found to be below 5.

**Results**
Our web application provides an intuitive and informative platform that enables users to access 20 years of SB-PFT results at both the school and district levels, create customized data maps, and conduct in-depth analysis utilizing our curated dataset.

The web interface is organized into two main components: the ***California School-Based Physical Fitness Testing Map*** and ***Custom Mapping***. The California School-Based Physical Fitness Testing Map consists of three key sections: 1) an introduction page with general information and tutorials, 2) an interactive map displaying school-level SB-PFT results, and 3) a map showcasing SB-PFT results at the school district level. The Custom Mapping component allows users to create their own visualizations through two sections: 4) a page for uploading custom maps, and 5) a page for generating custom map visualizations.

This section begins by presenting and detailing the web application we have developed, including both the California School-Based Physical Fitness Testing Map and the Custom Mapping components. Following this, a case study is provided to highlight our dataset as a readily accessible and analytical resource.

*California School-Based Physical Fitness Testing Map: Design Components 1-3.* The user-friendly interface of the map application simplifies the visualization and exploration of complex SB-PFT data and provides a powerful platform for education stakeholders to evaluate and compare physical fitness across California schools and school districts. A key feature of the SB-PFT school and school district map is the color-coded legend bar, which facilitates quick visual differentiation of school and school districts based on the percentage of students in the HFZ. The color gradient on the map serves as an indicator of student fitness levels: the bluer the area, the higher the percentage of students in the HFZ; conversely, the more orange the area, the lower the percentage. Schools and districts shaded in dark gray represent those for which data is not available. Users can tailor the data display through a dropdown menu and a multi-selection box, enabling a refined search by grade, test subject, school year, and specified geographic areas. The default map page on a school level, as shown in Figure 1, presents a comprehensive view of the percentage of 5th graders in HFZ on aerobic capacity assessment of each school in California for the 2018-2019 school year. Auto-clustering is enabled to dynamically group nearby points of interest to simplify navigation and enhance the readability of data at various zoom levels. When the map is zoomed out, smaller circles representing individual schools will cluster into a bigger circle with the number of schools on it. When users click the circle, the map will automatically zoom in and move the focus to areas surrounding the circle. The pop-up window is designed to provide a quick snapshot of essential information about a particular school. It displays the school name, address, the specific school district it belongs to, and the SB-PFT result of the selected school.

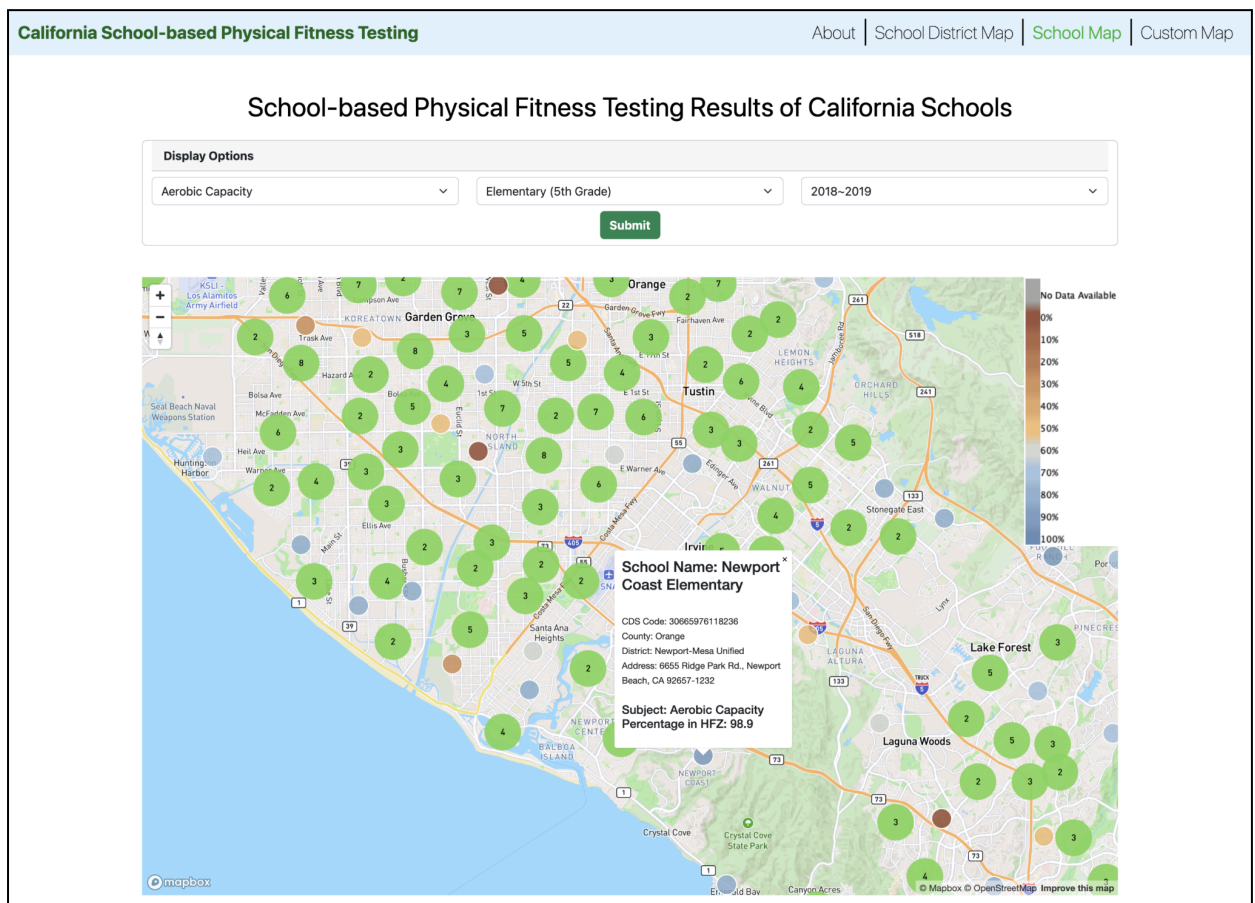

**Figure 1.** Interface of SB-PFT School Map Showing the Percentage of 5th Grade Students in the Healthy Fitness Zone (HFZ) for Aerobic Capacity Assessment in California, 2018-2019 Academic Year

Figure 2 presents the default visualization of the school districts map page, the percentage of 5th grade students in the HFZ in California in aerobic capacity during the 2018-2019 academic year. To offer detailed insights into regional fitness levels, users can click a district on the map and trigger a pop-up displaying the detailed names of county and school district, and the percentage. The pop-up window is configured to provide key information including the county in which the school district is located, the name of the school district, and the SB-PFT result of the selected school district.

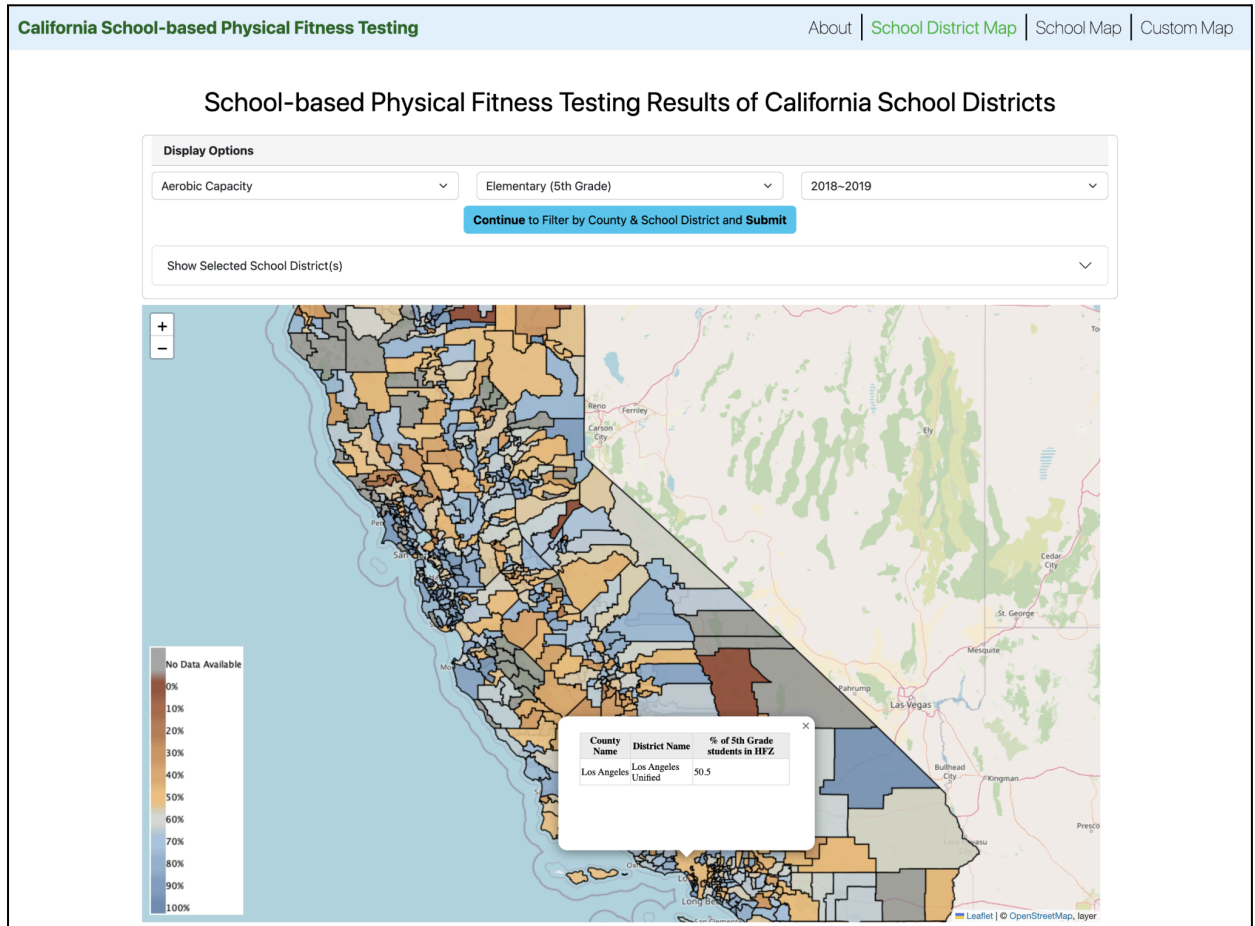

**Figure 2.** Interface of SB-PFT School District Map Showing the Percentage of 5th Grade Students in the Healthy Fitness Zone (HFZ) for Aerobic Capacity Assessment in California, 2018-2019 Academic Year

***Custom mapping: Design Components 4-5.*** Our web-based application allows users to access the app from a web browser without installing any additional software and uploading a dataset to create their own custom school district map. As depicted in Figure 3, we offer users the flexibility to explore and visualize school district-related data of California. The interface allows users to submit multiple files in .xlsx or .csv formats and assign unique names to each uploaded file for integration with the existing map data. To ensure the utility and efficiency of this tool, the feature includes user instructions detailing file format requirements, column naming guidelines for dataset integration, and file size requirement to prevent server overload. A demo video for the customized function is provided to enhance the analytical capabilities of our application. The customized mapping functionality is successfully tested using data on children's health insurance coverage across California school districts, derived from the American Community Survey – Education Children's Tabulation (ACS-ED) provided by the National Center for Education Statistics[39] (Figure 4).

**Figure 3.** Interface for Custom School District Mapping Creation: Instructions for Uploading a Custom Dataset

***Case Study - Integrated SB-PFT Results with SDoH Data for Explanatory Modeling.*** According to the results from our multiple linear regression models (see Table 1), we found that the percentage of non-institutionalized civilians with public health insurance, along with the mean family income within a school district, significantly influences the percentage of students in the HFZ for both aerobic capacity and body composition of a school district. As the proportion of non-institutionalized civilians with public health insurance increases by 1%, the percentage of students in HFZ decreases by 0.28% for aerobic capacity and by 0.15% for body composition of a district. Conversely, mean family income is positively correlated with the percentage of students in HFZ; a $10,000 increase in mean family income is associated with a 0.67% increase in the percentage of students in HFZ for aerobic capacity and a 0.61% increase for body composition. The percentage of the population speaking languages other than English at home is not a significant factor for the percentage of students in HFZ for aerobic capacity, however, every percent increase in this population is associated with a 0.14 decrease ($P<0.001$) in the percentage of students in HFZ for body composition. While the rates of housing units without a vehicle (estimate = -0.34, $P<0.05$) and households with a computer (estimate = -0.35, $P<0.001$) are both linked to a decrease in the rate of students in HFZ for aerobic capacity, they do not have a significant impact on the rate of students in HFZ for body composition. Our findings indicate that the socioeconomic conditions of a school district impact the physical fitness of children residing there.

The case study illustrates that our integrated SB-PFT results, with 20 years of school and school district level data, serve as a readily accessible resource. Users can download our preprocessed dataset, which can be effectively combined with external data sources for data analysis focused on children's fitness.

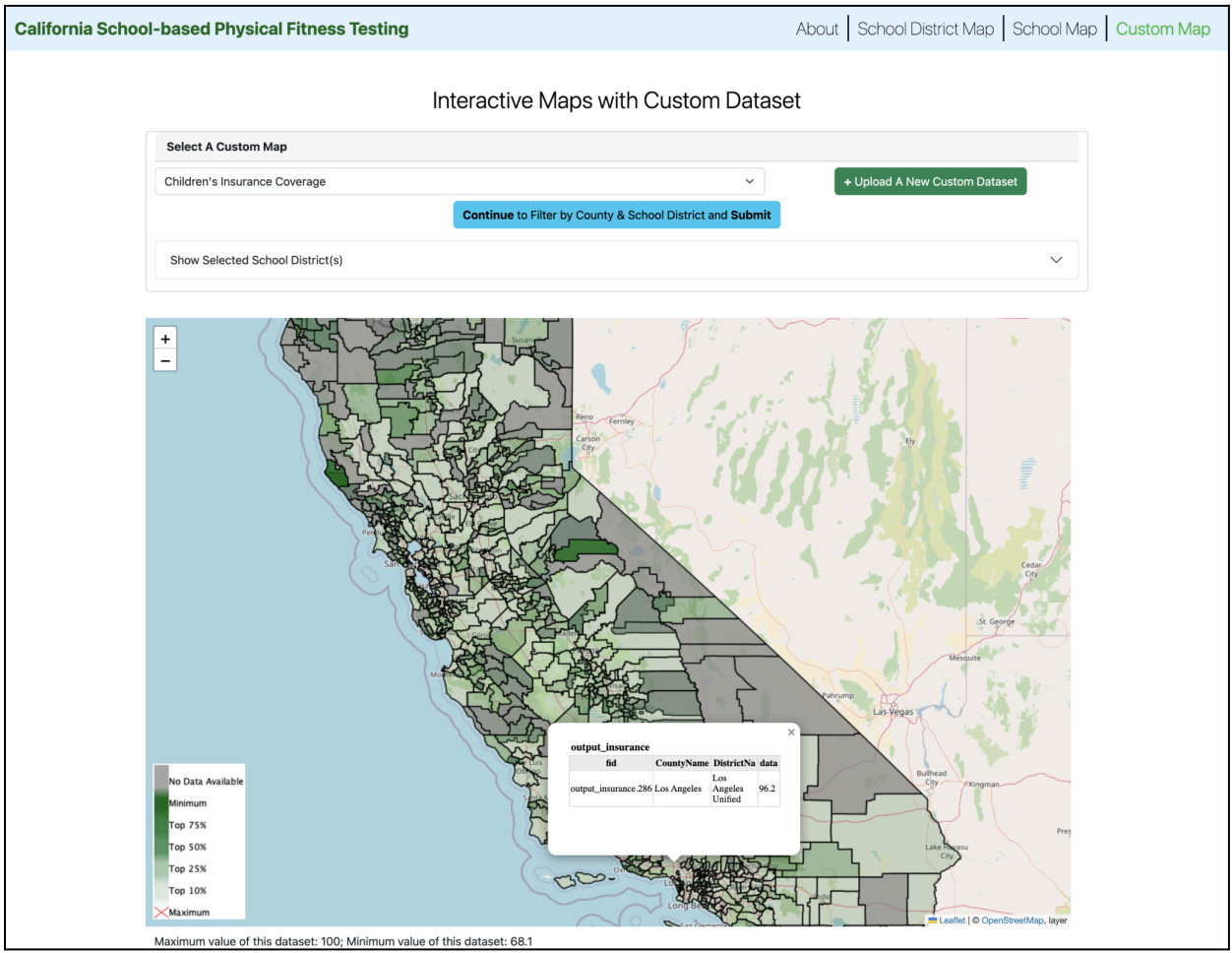

**Figure 4.** Custom Mapping Showcase: Mapping of Children's Health Insurance Coverage Across California School Districts Using ACS-ED Data

**Table 1.** Multi Linear Regression Model Results: Impact of SDoH on Children's Physical Fitness Indicators

|  | % of students in HFZ Aerobic Capacity | | % of students in HFZ Body Composition | |
|---|---|---|---|---|
| **Variable** | **Coef. Estimate** | **P-value** | **Coef. Estimate** | **P-value** |
| % of Population Speaking a Language Other than English at Home | -0.02336 | 0.251 | -0.14396 | ***<0.001 |
| % of Noninstitutionalized Civilians with Public Health Insurance | -0.27538 | ***<0.001 | -0.15155 | ***<0.001 |
| % of Households with a Computer | -0.34591 | ***<0.001 | -0.04178 | 0.437 |
| % of Occupied Housing Units with No Vehicles | -0.33619 | *0.0105 | -0.10991 | 0.214 |
| Mean Family Income (in 10000 dollars) | 0.67102 | ***<0.001 | 0.60278 | ***<0.001 |

**Discussion**
This study employed geospatial visualization techniques on SB-PFT data, revealing substantial disparities in physical fitness across schools and school districts in California. Our user-friendly customized function enables school district mapping creation and demonstrates the potential of data-driven geospatial analytics to uncover health disparities among school-aged children. We observed that certain districts predominantly show poor fitness levels, and our case study revealed possible links between SB-PFT results and various social factors. Additionally, mapping at the school level to highlight outliers—schools with higher fitness levels in generally low fitness areas—offers insights into effective health interventions for children and adolescents. This study highlights the benefits of geospatial visualization for understanding disparities in physical fitness among school-aged children and emphasizes the need for public health data to be more accessible and interpretable. Our findings support the use of geospatial mapping to explore how social and environmental factors affect children's physical fitness and engage the policy makers, community, school, and family to improve physical activity levels.

The case study demonstrates that SDoH at the school district level impacts the percentage of students falling within HFZ for aerobic capacity and body composition, with family income, public insurance coverage (such as Medicaid and Medicare) for non-institutionalized civilians, and household vehicle ownership being significant influencers. Prior research has shown that community level factors such as park access, walking distance, and socioeconomic disparities have an impact on children's body composition, diet, and physical activity [45, 46]. While results from our case study align with the existing body of knowledge, we are among the first to use geospatial visualization to study children's physical fitness at school district levels, linking it with SDoH information. These findings indicate potential targets for action to improve modifiable factors that may influence a students overall fitness and health development.

Future work includes integrating additional data to uncover the relationship between SDoH and the outcomes of SB-PFT. The future versions of this application will integrate more current datasets, capturing the resumption of the SB-PFT program of the COVID-19 pandemic and assess its ongoing impact. We aim to improve the visualization feature by integrating statistical analysis functionalities for visualizing trends and patterns over time, and enabling detailed maps for specific subgroups (racial-ethnic background, gender, socioeconomic status) for a comprehensive understanding of PFT trends across various sociodemographic factors. Further, we plan to enhance the utility and relevance of the application by incorporating SB-PFT data from other states, offering a national outlook that not only broadens the tool's use but also becomes a key asset for wider educational and policy development.

**Conclusion**
This study presents a novel web application to visualize and interpret SB-PFT data and reveal children's health and fitness disparities at both school and district levels across California. In addition, our designed custom mapping tool allows users to overlay data from other sources with the SB-PFT results and explore unique comparative perspectives. By analyzing the SB-PFT results with socioeconomic data from the American Community Survey, our case study not only highlights the potential of integrating our SB-PFT results with other datasets to gain insights into children's physical fitness but also reveals the relationship between social determinants, especially financial-related factors, and children's fitness indicators. Future work will focus on providing SB-PFT visualization for racial-ethnic and socioeconomic subgroups visualizations of SB-PFT, and supporting statistical features for longitudinal data comparisons. This application is a resource for broad stakeholders to assess children's fitness and supports researchers in conducting detailed analyses with social determinants of health. Our study highlights the importance of data-driven methods in shaping health and education policies, advocating for the use of technology to enhance children's physical fitness and overall health.

**Useful Links**
SB-PFT Interactive Map Home Page:
  http://ec2-54-176-149-48.us-west-1.compute.amazonaws.com:3000/
Dataset Repository, Web App User Guide, and Local Setup Instructions:
  https://github.com/kaiyuanh2/physicalmapapp